\documentclass[
 reprint,
nofootinbib,
 amsmath,amssymb,
 aps,
floatfix 
]{revtex4-1}

\usepackage{graphicx}
\usepackage{dcolumn}
\usepackage{bm}
\usepackage[colorinlistoftodos]{todonotes}
\usepackage[export]{adjustbox}%
\usepackage{natbib}

\usepackage{xcolor}
\colorlet{rcolor}{black}  

\newcommand{\dropcap}[1]{#1}
\newcommand{\captionof}[2]{\caption{#2}}

\DeclareMathOperator*{\argmax}{arg\,max}

\begin{document}

\title{Why {\em ex post} peer review encourages high-risk research\\while {\em ex ante} review discourages it}

\author{Kevin Gross}
\email{krgross@ncsu.edu}
\affiliation{Department of Statistics \\ North Carolina State University \\ Raleigh, NC USA}

\author{Carl T. Bergstrom}
\email{cbergst@u.washington.edu}
\affiliation{Department of Biology \\ University of Washington \\ Seattle, WA USA} 

\date{\today}

\begin{abstract}
Peer review is an integral component of contemporary science.  While peer review focuses attention on promising and interesting science, it also encourages scientists to pursue some questions at the expense of others.  Here, we use ideas from forecasting assessment to examine how two modes of peer review --- {\em ex ante} review of proposals for future work and {\em ex post} review of completed science --- motivate scientists to favor some questions instead of others.  Our main result is that {\em ex ante} and {\em ex post} peer review push investigators toward distinct sets of scientific questions.  This tension arises because {\em ex post} review allows an investigator to leverage her own scientific beliefs to generate results that others will find surprising, whereas {\em ex ante} review does not.  Moreover, {\em ex ante} review will favor different research questions depending on whether reviewers rank proposals in anticipation of changes to their own personal beliefs, or to the beliefs of their peers. The tension between {\em ex ante} and {\em ex post} review puts investigators in a bind, because most researchers need to find projects that will survive both.  By unpacking the tension between these two modes of review, we can understand how they shape the landscape of science and how changes to peer review might shift scientific activity in unforeseen directions.
\end{abstract}

\maketitle

\begin{figure}
  \begin{center}
    \includegraphics[width=\linewidth]{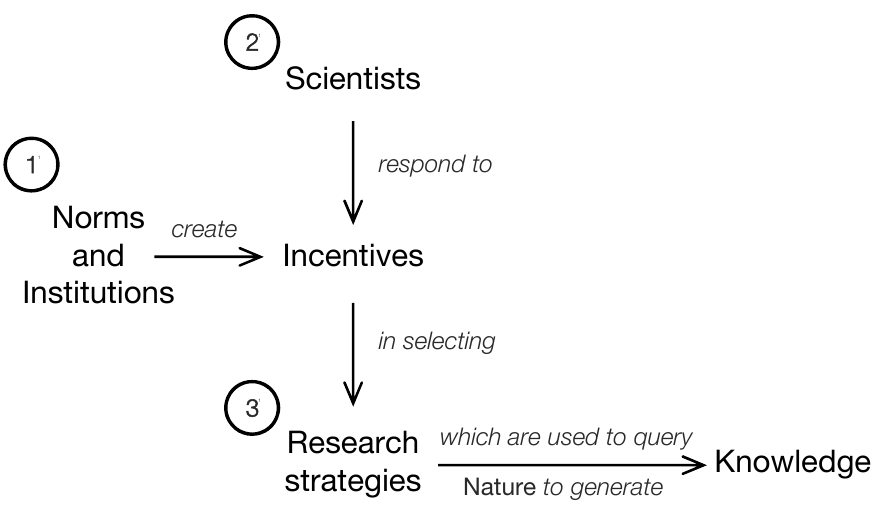}
  \end{center}
  \captionof{figure}{\textbf{Scientific norms and institutions create incentives that shape the knowledge generated by scientific research.} (1) Scientific institutions such as funding agencies, journals, scholarly societies, and university departments interact with scientific norms such as citation practices and the priority rule for credit to create a suite of incentives for scientific researchers.  (2) Scientists are motivated by curiosity, prestige, financial reward, and a desire to continue working within the research community, and  shape their practices accordingly. (3) Through these research practices, scientists query nature to generate a collection of knowledge --- correct or otherwise --- about how the natural world works.}
  \label{fig:viewpoint}
\end{figure}

\dropcap{O}ur understanding of the natural world is shaped by the decisions scientists make about what systems to study, what models to compare, what hypotheses to test, and other aspects of their research strategies. While investigators may be driven by intrinsic curiosity, they are also constrained by the realities of the scientific ecosystem in which they operate, and motivated by the other incentives they confront \citep{arrow1962economic,kaplan1965norms,dasgupta1987simple,hull1988science,kitcher1993advancement,brock1999formal,strevens2003,glaeser2006,zollman2009optimal,kleinberg2011mechanisms,boyer2014bird}. Therefore the choices that researchers make are not purely epistemic, but are also {\color{rcolor} influenced} by the norms and institutions that govern and support the scientific endeavor. These norms and institutions thus shape the concordance between our knowledge of nature and its actual workings (Fig.~\ref{fig:viewpoint}).  While this observation may sound pessimistic, it affords us an opportunity: if we can understand the epistemic consequences of the social structures within which science is conducted, we may be able to nudge the scientific ecosystem in directions that produce more reliable and expansive knowledge about the world.  


In this article, we examine how the institution of peer review shapes the landscape of science by pushing scientists towards some questions and away from others.  One purpose of peer review --- perhaps it's nominal purpose --- is to ensure quality and rigor in science \citep{polanyi1962republic,merton1973sociology,jefferson2002effects,bornmann2011scientific}.  Another is to improve the quality of research design and communication \citep{goodman1994manuscript}. However, peer review also serves as a filter that focuses effort and attention on promising questions and interesting results \citep{horrobin1990philosophical,kostoff2004research,baldwin2018scientific,avin2019centralized}.  Yet by screening out certain projects and results in favor of others, peer review also creates incentives that investigators must confront when structuring their research program.  As a result, peer-review filters do more than merely screen out weak science: they shape the kinds of questions that researchers set out to study in the first place.  

Moreover, most research must navigate two separate rounds of peer review before reaching fruition. {\em Ex ante} peer review occurs before the work is conducted and before the results are known, as happens when scientists submit grant proposals. {\em Ex post} peer review takes place after the work is conducted and the results have been determined, as happens when scientists submit a manuscript to a scientific journal or juried conference.  When deciding what problem to work on, researchers need to consider both whether the work is likely to be funded, and whether the anticipated results of the work will be publishable. 

From this perspective, how can we understand the process by which scientists select research questions?  We follow the foundations laid by the ``new economics of science'' \citep{partha1994toward,stephan1996economics}, envisioning scientists as rational actors seeking to achieve epistemically desirable states \citep{kitcher1990}.  Our purpose is not to cast scientists as automatons or to discount individual curiosity but instead to recognize that scientists are, to borrow Philip Kitcher's memorable phrase, ``epistemically sullied'' \citep{kitcher1990}.  That is, in addition to following their own curiosities, they also want to do work that is rewarded.   In academia, all else equal, work is rewarded when it has scientific ``impact'', that is, when it shifts researchers' understanding of the natural world.

We build from the premise that work is valued when it changes the beliefs of the scientific community \cite{davis1971thats,goldman1991economic,frankel2021findings}. Yet, {\em ex ante}, reviewers can only make educated guesses about how proposed research will change their beliefs, because they don't know the outcome of the experiments. {\em Ex post}, reviewers have the results in hand and can assess their value directly. Does this distinction matter? Do {\em ex ante} and {\em ex post} peer review encourage the same science?  If not, how and why do they differ?

In this article, we analyze how {\em ex ante} and {\em ex post} peer review influence investigators' research programs and thereby shape the scientific landscape.  Our approach is positive, not normative, as we describe how {\em ex ante} and {\em ex post} peer review shape research programs but are not yet equipped to evaluate whether this {\color{rcolor} process} is good or bad for science. We argue that understanding how {\em ex ante} and {\em ex post} peer review shape investigators' research yields a variety of useful insights.  Among its other implications, this understanding illuminates why research proposals often feature low-risk science, suggests how funding competitions might be structured to encourage more ambitious proposals, and indicates that recent momentum toward {\em ex ante} review of journal publications \citep[e.g.][]{nosek2014method, nosek2018preregistration, chambers2020frontloading} may {\color{rcolor} redirect} scientific activity in previously unanticipated ways.  We first {\color{rcolor} develop} the key ideas verbally, follow with our formal model, and then present a numerical example to illustrate the results.

\section*{Theory}

\subsection*{Evaluating experiments by anticipated shift in belief}

We envision a population of investigators who use data to arbitrate among a collection of competing epistemic objects.  These objects may be theories, models, hypotheses, effect sizes, or best practices.  Each investigator has a private set of beliefs that captures what she knows or suspects about the objects in question.  
Beliefs may differ among investigators, and if they do, each investigator has a working understanding of how beliefs vary among her peers. 

We assume that scientists seek to truthfully discriminate among the competing epistemic objects they consider, and to act accordingly.  As such, we assume that investigators value experiments and results insofar as they shift either their own scientific beliefs or the beliefs of others. Scientific beliefs can shift in a number of ways: by moving support among competing theories, by breaking open new lines of inquiry, by consolidating previously disparate ideas, or by replicating an experiment to demonstrate its external validity, 
to name a few. Shifts in belief are not limited to overturning the prevailing hypothesis; beliefs also shift when confidence in a leading idea or theory grows.  In contrast, studies that leave scientific beliefs unchanged have little value to the scientific community.  

To sketch the basic idea of our model, suppose that an investigator contemplates an experiment that would help her and/or her peers arbitrate among several competing epistemic objects.  Once an experiment is completed, any observer who learns of the outcome updates their beliefs using Bayesian reasoning.  As we explain in more detail below, we can quantify the value of this belief shift as the amount by which the observer perceives it to have improved her utility in a decision problem that rewards accurate beliefs \citep{frankel2019quantifying}. The anticipated value of an experiment is then determined by weighting the value of the possible outcomes and their corresponding shifts in belief by the perceived probabilities of obtaining those outcomes.  In other words, the anticipated value of an experiment is

\begin{multline}
    \text{value of experiment} = \\ \sum_{y \ \in \ \text{outcomes}} \Big\{ \left(\text{probability of outcome } y\right) \times \\ \left(\text{value of shift in beliefs from outcome } y\right) \Big\}. 
    \label{eq:ex-ante-word}
    \end{multline}

Our central observation is that there are several different ways that an investigator might value a potential experiment, based on (a)  whose beliefs are used to {\color{rcolor} assign probabilities to the} experiment's potential outcomes and (b) whose beliefs will shift upon observing the outcome.  In the simplest case, an epistemically unsullied investigator uses her own beliefs to weight the possible experimental outcomes and cares only about subsequent shifts in her own beliefs.  This investigator values experiments as expressed in eq.~\ref{eq:investigator-private-word}. 
\begin{figure*}[bt!]
\begin{multline}
    \text{private value of experiment to an investigator} = \\ \sum_{y \ \in \ \text{outcomes}} \Big\{ \left(\text{{\color{blue} investigator's} perceived probability of obtaining outcome } y\right)  \times \\ \left(\text{value of shift in {\color{blue} investigator's} belief from outcome } y\right) \Big\}, 
    \label{eq:investigator-private-word}
    \end{multline}
\end{figure*} 
We call this the private value of the experiment to an investigator.

In contrast, an investigator who cares most about publishing her results in a high-profile journal wishes instead to shift the beliefs of her peers, because peer reviewers will only recommend publication if they perceive the outcome as having scientific value to the broader scholarly community.  However, the investigator continues to weight the potential outcomes by her own beliefs.  Thus, this investigator values experiments that she anticipates will shift her peers' beliefs according to eq.~\ref{eq:investigator-public-word}.
\begin{figure*}[bt!]
\begin{multline}
    \text{public value of experiment to an investigator} = \\ \sum_{y \ \in \ \text{outcomes}} \Big\{ \left(\text{{\color{blue} investigator's} perceived probability of outcome } y\right) \times \\ \left(\text{value of shift in {\color{red} peers'} beliefs from outcome } y\right) \Big\},
    \label{eq:investigator-public-word}
\end{multline}
\end{figure*} 
We call this the public value of the experiment to an investigator.

Separately, an investigator who needs to obtain a grant to conduct her experiment (or needs to have the experiment blessed by her peers in some other way)  must convince reviewers of the value of the experiment before the outcome is known.  The formulation of the experiment's value here is more subtle, and leads to two possible definitions.  In both definitions, peer reviewers use their own beliefs to assess the probabilities of various experimental outcomes.  However, we can obtain two different definitions depending on {\em whose} beliefs the peers think about shifting.  In one definition, we might think about peers who only care about shifting their own beliefs, in which case the value of the experiment is given by eq.~\ref{eq:reviewer-private-word}.  
\begin{figure*}[bt!]
\begin{multline}
    \text{private value of experiment to proposal reviewers} = \\ \sum_{y \ \in \ \text{outcomes}} \Big\{ \left(\text{{\color{red} peers'} perceived probability of outcome } y\right) \times \\ \left(\text{{value of shift that \color{red} peers} expect in {\color{purple} their own beliefs} from outcome } y\right) \Big\},
    \label{eq:reviewer-private-word}
\end{multline}
\end{figure*} 
On the other hand, the peer reviewers might evaluate the experiment with respect to how they anticipate the experiment will shift the beliefs of the community (that is, their peers).  In this case the value of the experiment is given by eq.~\ref{eq:reviewer-public-word}.  
\begin{figure*}[bt!]
\begin{multline}
    \text{public value of experiment to proposal reviewers} = \\ \sum_{y \ \in \ \text{outcomes}}  \Big\{ \left(\text{{\color{red} peers'} perceived probability of outcome } y\right) \times \\ \left(\text{value of shift that {\color{red} peers} expect in {\color{purple} other peers' beliefs} from outcome } y\right) \Big\}.
    \label{eq:reviewer-public-word}
\end{multline}
\end{figure*} 
We call eq.~\ref{eq:reviewer-private-word} the private value of the experiment to proposal reviewers, and call eq.~\ref{eq:reviewer-public-word} the public value of the experiment to proposal reviewers.  While the distinction between eq.~\ref{eq:reviewer-private-word} and eq.~\ref{eq:reviewer-public-word} may seem subtle, we will see below that it can have a large impact on the types of proposals that review panels favor. 

To illustrate how these criteria differ, consider an investigator who has the opportunity to definitively determine whether or not life exists on Mars.  Further, suppose that everyone in the relevant scholarly field (including the investigator) is convinced almost beyond doubt that they already know whether or not Martian life exists.  However, these researchers disagree --- 70\% of the community is convinced that Martian life does not exist, while the other 30\% is equally well convinced of the opposite.  Finally, suppose that upon learning the outcome of this definitive experiment, researchers who are surprised by the outcome and are forced to change their beliefs value their learning at 1, while researchers who are unsurprised by the outcome value their (non-)learning at 0.

How eager is the investigator to pursue this experiment?  If she is driven only by her own curiosity, then she has no interest in doing the experiment and privately values it at 0 (eq.~\ref{eq:investigator-private-word}), because she is convinced that the experiment will merely confirm what she already knows.  Her effort is better spent elsewhere.  If the investigator wants to publish her result, then she will value the experiment based on what she expects the community to learn from its outcome (eq.~\ref{eq:investigator-public-word}).  If she is convinced that life does not exist on Mars, then she determines the public value of the experiment to be 0.3, because she expects that, upon observing that Martian life definitively does not exist, 30\% of her peers will change their beliefs and value the discovery at 1, while 70\% of her peers will not shift their beliefs and value the discovery at 0.  (Note that this calculation holds regardless of whether her peers value the outcome based on their own private belief shift or based on the public belief shift of the community.  The average private belief shift among her peers is 0.3, and everyone also agrees that the public belief shift is 0.3.)  Similarly, if the focal investigator is convinced that life does exist on Mars, she determines the public value of the experiment to be 0.7.

Now suppose that the investigator has to convince her peers to fund her experiment before it can be conducted.  If her peers are only interested in their own private learning (eq.~\ref{eq:reviewer-private-word}), then the experiment will have no value to the community, because everyone in the community expects the experiment to confirm their own beliefs.  On the other hand, if her peers are interested in shifting the community's beliefs, then 70\% of the community (those who don't belief in Martian life) will value the experiment at 0.3, and 30\% of the community will value the experiment at 0.7.  Thus the average valuation across a collection of peers is (0.7)(0.3) + (0.3)(0.7) = 0.42 (eq.~\ref{eq:reviewer-public-word}). As this example illustrates, the perceived value of an experiment can differ dramatically based on which of the above criteria is used. 

{\color{rcolor} Overall, scientists are left in a bind}. Most researchers must obtain grants {\em and} publish.  They face an inescapable tension between those experiments that are most likely to win reviewers' favor when being proposed, versus those experiments that are most likely to generate outcomes that win referees' favor when submitted for publication.  We suspect that this tension will be familiar to anyone who has shepherded a research idea through the cycle from grant proposal to journal publication.

\subsection*{Formal model}

In this section, we derive mathematical expressions for eqs.~\ref{eq:investigator-private-word}--\ref{eq:reviewer-public-word}.  The central task is to quantify the value of the belief shift that results from observing a particular experimental outcome; all the rest simply involves {\color{rcolor} integrating} over suitable {\color{rcolor} distributions}.  We defer that central task for the moment and begin with some preliminaries. 

Let $\mathcal{X}$ denote a set of competing epistemic objects, such as theories, claims, or parameter values, and let $x \in \mathcal{X}$ denote a member of $\mathcal{X}$.  We assume that exactly one member of $\mathcal{X}$ obtains in nature, that is, it is the ``true'' value. Let $X$ be a random variable that represents this true but unknown state of nature.  An investigator's beliefs across the objects in $\mathcal{X}$ are captured as a probability distribution on $\mathcal{X}$.  Let {\color{rcolor} $\mathcal{P}$} denote the set of such probability distributions, and use ${\color{rcolor} P, Q, R \in \mathcal{P}}$ to denote possible beliefs.  
{\color{rcolor} Finally, let $\Pi$ be a distribution across $\mathcal{P}$ that represents the distribution of beliefs across (a unit mass of) investigators in the community.}  

Investigators conduct experiments to learn about the value of $X$.  Let $Y$ be a random variable that denotes the outcome of an experiment designed to reveal something about $X$, and let $y$ be a possible value of $Y$.   Let {\color{rcolor} $F(y|x)$} give the conditional probability distribution of $y$ when $x$ is the state of nature that prevails.  We assume that all investigators agree upon the probability structure in {\color{rcolor} $F(y|x)$} despite their differing beliefs about $X$.  Because {\color{rcolor} $F(y|x)$} represents an agreed-upon procedure for interpreting the evidence, we  might identify {\color{rcolor} $F(y|x)$}  as the scientific paradigm.  We assume that any investigator who begins with prior belief {\color{rcolor} $P$} and observes $Y=y$ uses Bayesian reasoning to update their beliefs to a resulting posterior belief {\color{rcolor} $Q(P,y) \in \mathcal{P}$}.

We now turn to quantifying the value of a belief shift.  Here, we focus on the specific case where {\color{rcolor} each} researcher's utility is determined by how accurately her beliefs anticipate the state of nature, or observable phenomena that derive from it.  However, predictive validity is not the only conceivable measure of scientific progress,  
{\color{rcolor} and maximizing such is not the only conceivable aim that researchers might have.} 
In the appendix, we use the results of Frankel \& Kamenica \citep{frankel2019quantifying} to extend our approach to any decision problem that researchers may face that rewards accurate beliefs, such as developing new technologies or informing policy. 
Hence this framework is flexible enough to encompass a wide range of settings, both basic and applied. 

To {\color{rcolor} address} the specific case of researchers seeking accurate beliefs, we turn to the well-developed theory of scoring rules from the field of forecasting assessment \citep{brier1950verification,brown1970probabilistic,gneiting2007strictly, brocker2009reliability}.  A scoring rule {\color{rcolor} $S(P,x)$} is a function that assigns a numeric value to each pair of probability distributions {\color{rcolor} $ P \in  \mathcal{P}$} and objects $x \in \mathcal{X}$. 
{\color{rcolor} In forecasting assessment, $S(P,x)$ measures the surprise that results when a the forecast {\color{rcolor}$P$} is issued for $X$, and $x$ is subsequently observed.  Here, we follow the convention that larger scores indicate an outcome that is less consistent with the forecast \citep{brocker2009reliability}.}
%
{\color{rcolor} We restrict our attention to strictly proper scoring rules, under which} a forecaster optimizes her expected score by issuing a forecast that reflects her true beliefs \cite{brown1970probabilistic}.   
{\color{rcolor} Thus, strictly proper} scoring rules elicit honest forecasts.  In {\color{rcolor} the present} context, if an investigator holds beliefs {\color{rcolor}$P$ for $X$}, and the true state of nature is revealed to be $x$, then the score {\color{rcolor} $S(P,x)$} quantifies how well or poorly the investigator's beliefs anticipate $x$.  Scoring rules are thus useful for quantifying scientific progress because they reward movement towards beliefs that successfully anticipate observable phenomena while also rewarding faithful awareness of uncertainty. Both are key virtues in science.  Scoring rules also generalize some familiar notions from information theory \citep{cover2012}, as we will see below. 

Equipped with a scoring rule, the divergence from belief {\color{rcolor} $P_2$} to {\color{rcolor} $P_1$} is defined as
\begin{equation}
{\color{rcolor} d(P_2 || P_1) = \int \left( S(P_1, x) - S(P_2, x) \right) \, dP_2(x)}.
\label{eq:divergence}
\end{equation}
The divergence ${\color{rcolor} d(P_2 || P_1)}$ measures the loss in predictive fidelity that results from using {\color{rcolor} $P_1$} to forecast $X$, when in fact $X$ is actually distributed as {\color{rcolor} $P_2$}. 
A proper scoring rule is one for which the divergence between any two beliefs is non-negative, and a strictly proper scoring rule is a proper scoring rule with the added condition that ${\color{rcolor} d(P_2 || P_1) = 0}$ if and only if {\color{rcolor} $P_1 = P_2$} \citep{gneiting2007strictly, brocker2009reliability}.  
{\color{rcolor} We define the value of outcome $y$ to an investigator who holds prior beliefs $P$ as the divergence from the investigator's posterior beliefs to their prior beliefs, $d(Q(P,y) || P)$.}  In other words, the value of experimental outcome $y$ is the amount by which an observer thinks that her forecast of $X$ has improved by observing $y$ and updating her beliefs accordingly. {\color{rcolor}   We denote this value by $v(y,P)$.}

{\color{rcolor} To push the notation a bit further, let $F(y;P) = \int \, F(y|x) \, dP(x)$ give the distribution of $Y$ anticipated by an investigator who believes $P$. Let $F(y) = \int \, F(y; P) \, d\Pi(P)$ give the anticipated distribution of $Y$ averaged across the researchers' beliefs in the community, and let $v(y) = \int \, v(y; P) \, d\Pi(P)$ give the average value of outcome $y$ to the community.}  We can now write  expressions corresponding to eqs.~\ref{eq:investigator-private-word}--\ref{eq:reviewer-public-word}.   For eq.~\ref{eq:investigator-private-word}, an epistemically unsullied investigator who believes {\color{rcolor} $P$} values an experiment by her expected shift in her own beliefs, which is {\color{rcolor} written as}
\begin{equation}
{\color{rcolor} \int \, v(y,P) \, dF(y;P)}.
\label{eq:investigator-private}
\end{equation}  
For eq.~\ref{eq:investigator-public-word}, an investigator {\color{rcolor} who believes $P$ and is} most interested in publishing her results values her experiment based on how much she expects the outcome to shift the beliefs of {\color{rcolor} the community, which is given by}
\begin{equation}
{\color{rcolor}\int \! \! \int \, v(y,R) \, d\Pi(R) \, dF(y;P) = \int \, v(y) \, dF(y;P)}.
\label{eq:investigator-public}
\end{equation}
Eq.~\ref{eq:reviewer-private-word} is the average amount by which a reviewer expects the experiment to shift her own private beliefs, which can be written as {\color{rcolor} eq. \ref{eq:investigator-private} integrated over the distribution of peer reviewers' private beliefs:} 
\begin{equation}
    {\color{rcolor}\int \! \! \int \, v(y,P) \, dF(y;P) \, d\Pi(P)}.
    \label{eq:reviewer-private}
\end{equation}
Finally, eq.~\ref{eq:reviewer-public-word} is the average amount by which a reviewer expects the experiment to shift the beliefs of a the community, which {\color{rcolor} is given by eq.~\ref{eq:investigator-public} integrated over the distribution of reviewers' private beliefs:}
\begin{equation}
    {\color{rcolor} \int \! \! \int \, v(y) \, dF(y;P) \, d\Pi(P) = \int \, v(y) \, dF(y)}.
    \label{eq:reviewer-public}
\end{equation}

Before moving on, we note a few useful connections with information theory \citep{cover2012}.  Suppose $\mathcal{X}$ is discrete, and let $p_x$ denote the probability {\color{rcolor} that the distribution $P$ assigns} to $x \in \mathcal{X}$.  Suppose we use the  ``ignorance'' score $S({\color{rcolor} P}, x) = -\log_2 p_x$, also known as the surprisal.
Plugging the ignorance score into eq.~\ref{eq:divergence} yields the familiar Kullback-Leibler divergence.  Further, the expected belief shift to an investigator who believes {\color{rcolor} $P$} (eq.~\ref{eq:investigator-private}) becomes the mutual information between the investigator's beliefs and the experiment $Y$.  In the same vein, eq.~\ref{eq:reviewer-private} becomes the average mutual information between $Y$ and the beliefs of a randomly chosen reviewer.

\section*{Numerical example}

The following example illustrates how different criteria for evaluating experiments pull investigators in conflicting directions.   {\color{rcolor} In this example, investigators consider pursuing one of several possible research questions, each of which entails arbitrating between two competing claims.}  
For each such question, let $X=0$ code for one claim while $X=1$ codes for its competitor.  
An investigator's beliefs are captured by the probability that she assigns to the claim coded by $X=1$, which we denote by $p \in [0, 1]$.  ({\color{rcolor} In the notation of the formal model above, $p$  fully specifies a belief $P$.})  For a scoring rule, we use the Brier score $S({\color{rcolor} P}, x) = (x - p)^2$ \cite{brier1950verification}.

Each possible study {\color{rcolor} generates an outcome} $Y$, where $Y$ takes a Gaussian distribution with mean $\mu_0$ when $X=0$ and $\mu_1 \neq \mu_0$ when $X=1$.  The standard deviation of $Y$ is $\sigma_Y = | \mu_0 - \mu_1 | / 2$ regardless of whether $X=0$ or $X=1$.  Our qualitative results that follow are not sensitive to the choice of $\sigma_Y$, but they do reflect the assumption that $\sigma_Y$ is the same for both $X=0$ and $X=1$.  {\color{rcolor} Computations are conducted using R \cite{R2021}.}

\begin{figure}
  \begin{center}
    \includegraphics[width=\linewidth]{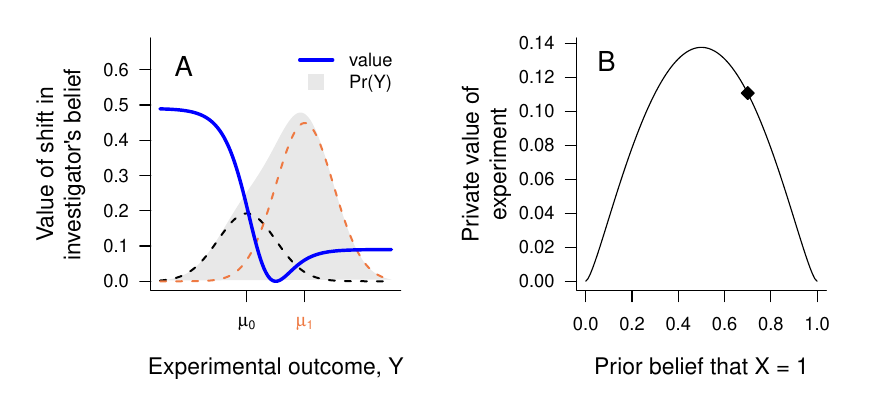}
  \end{center}
  \captionof{figure}{\textbf{An epistemically unsullied investigator values an experiment based on how much she expects the outcome will shift her own beliefs.}  (A) For the example described in text, suppose an investigator's prior belief that $X=1$ is $p=0.7$. The dashed lines show the distribution of the experiment's outcome $Y$ if $X=0$ ({\color{rcolor} black}) or $X=1$ ({\color{rcolor} orange}) with relative probabilities weighted by the investigator's prior beliefs. The shaded area gives the overall anticipated distribution of outcomes. The solid {\color{rcolor} blue} curve indicates how much the investigator would value each outcome, as quantified by the amount it would shift her beliefs.  Integrating this value curve over the shaded probability distribution gives her private value of the experiment. (B) The private value of the experiment as a function of the investigator's prior beliefs.  In this particular example, the investigator most favors studies for which her prior beliefs give equal credence to the two competing claims. The diamond shows the private value for the scenario in panel A.}
  \label{fig:unsullied-schematic}
\end{figure}

First consider an investigator who is most interested in satisfying her own curiosities 
(eq.~\ref{eq:investigator-private-word}).  While this investigator may allow for the possibility of a result that surprises her, she never expects such a result; instead, if she favors one claim over the other, then she expects that the experiment will confirm her beliefs more often than not (Fig.~\ref{fig:unsullied-schematic}A).  Because of the symmetry in this particular example, the investigator expects an experiment to be most informative to her when her prior beliefs are most uncertain, and thus she favors studies for which she gives equal credence to either claim ($p=0.5$, Fig.~\ref{fig:unsullied-schematic}B). 

\begin{figure}
  \begin{center}
    \includegraphics[width=\linewidth]{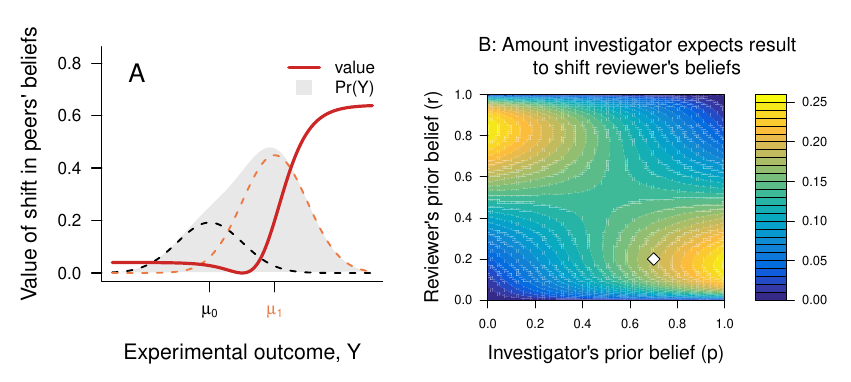}
  \end{center}
    \captionof{figure}{\textbf{An epistemically sullied investigator values an experiment based on how much she thinks the {\color{rcolor} outcome} will shift her peers' beliefs.} (A) As in Figure \ref{fig:unsullied-schematic}A, the investigator anticipates a distribution of experimental {\color{rcolor} outcomes (dashed lines and shaded region) based on her belief that $X=1$ with probability $p=0.7$.} However, the value of any possible outcome is now determined by how much it would the shift in beliefs of a peer reviewer. Here, the {\color{rcolor} solid} red curve illustrates the value to a reviewer who believes that $X=1$ with probability $r=0.2$. (B) The anticipated value of the experiment now depends on both the investigator's own beliefs and the reviewer's beliefs. The investigator favors experiments that maximize her chance of shifting her peers' beliefs substantially, and she does this by leveraging the differences between her beliefs and those of the peers.  The white diamond locates the setting in panel A. 
    }
    \label{fig:sullied-schematic}
\end{figure}

Next consider an investigator who is most interested in publishing her results (Fig.~\ref{fig:sullied-schematic}).  To simplify matters, suppose this investigator is a ``lone wolf''  in the sense that, {\color{rcolor} for any possible question, she believes that one of the two competing claims is true with probability $p$, while everyone else in the community agrees that this claim is true with probability $r$.  The investigator now considers questions that differ in their values of $p$ and $r$.  She anticipates a} study's outcome using her own beliefs, but values the outcome based on how much it would shift {\color{rcolor} her peers' beliefs} (Fig.~\ref{fig:sullied-schematic}A).  This investigator can leverage the difference between her beliefs and {\color{rcolor} her peers' beliefs} by conducting an experiment that she thinks will surprise them.  She favors studies where her beliefs are strongly at odds with the community, and she most prefers a study where she is absolutely convinced that one of the claims is true ($p=0$ or $p=1$), and her peers are largely --- though not wholly --- convinced that it is false (Fig.~\ref{fig:sullied-schematic}B).  

Notably, this investigator does not simply seek to maximize the difference between her beliefs and those of her peers (Fig.~\ref{fig:sullied-schematic}B).  Instead, she must account for the fact that if the community's beliefs are too strongly held ($r \approx 0$ or $r \approx 1$), it is unlikely that the experiment will generate a sufficiently compelling result to shift the community's beliefs far, {\color{rcolor} because the community will dismiss a surprising result as an experimental fluke \citep{frankel2021which}.}  Thus the investigator shies away from pursuing claims where the community considers the matter to be largely settled, no matter how strongly she may disagree with the consensus view.  This potentially explains why inquiry in a question ceases once a consensus emerges that a question is mostly settled, even if that consensus admits room for some reasonable doubt \citep{zollman2010epistemic, nissen2016publication}.

Now consider {\em ex ante} peer review of a proposed study. In the lone-wolf example, everyone other than the investigator has identical beliefs, so the distinction between whether reviewers evaluate the proposed experiment based on the change in their own private beliefs (eq.~\ref{eq:reviewer-private-word}) or the community's aggregate beliefs (eq.~\ref{eq:reviewer-public-word}) vanishes.  In either case, the peer reviewers would use their own beliefs both to assign probabilities to the experiment's possible outcomes and to determine the value of those outcomes.  Thus, the investigator has no opportunity to leverage her own private beliefs.  Instead, she must instead gravitate towards studies that her peers deem most informative --- which in this example are those for which $r = 0.5$ --- regardless of what she herself believes.

\subsection*{Peer reviewers with heterogeneous beliefs}

While the ``lone-wolf'' example is illuminating, it suffers from two shortcomings.  First, {\color{rcolor} real} investigators are members of the community whose beliefs they seek to change, and so questions where an investigator's views diverge strongly from those of the community are likely to be in short supply.  Second, we need a richer model of how beliefs vary among the community than to merely assert that everyone other than the focal investigator holds the same belief.  To see past these limitations, we conducted a small simulation.  This simulation envisions that investigators have access to a common, large pool of potential questions.  Questions differ only with regard to the distribution of beliefs across the community.  In the simulation, each investigator first samples a handful of candidate questions.  For each sampled question, the investigator is randomly assigned a belief from the distribution of beliefs attached to that question.  Because the investigator's belief is assigned in this way, the investigator's beliefs will accord with most of her peers' beliefs most of the time.  The investigator then pursues the question that is most promising based on one of the three criteria in eqs.~\ref{eq:investigator-public-word}--\ref{eq:reviewer-public-word}.  By repeating this process for many such investigators, we build up a collection of questions that reflect the scientific activity within this simulated community.  

To model the distribution of community beliefs, suppose that for each question the community is divided into a majority camp and a minority camp.  The proportion of investigators in the majority camp is drawn from a uniform distribution on $[0.5, 1]$.  Within each camp, every investigator holds the same beliefs.  Common beliefs within the majority and minority camps are drawn independently from a uniform distribution on $[0, 1]$.  

\begin{figure}
  \begin{center}
    \includegraphics[width=\linewidth]{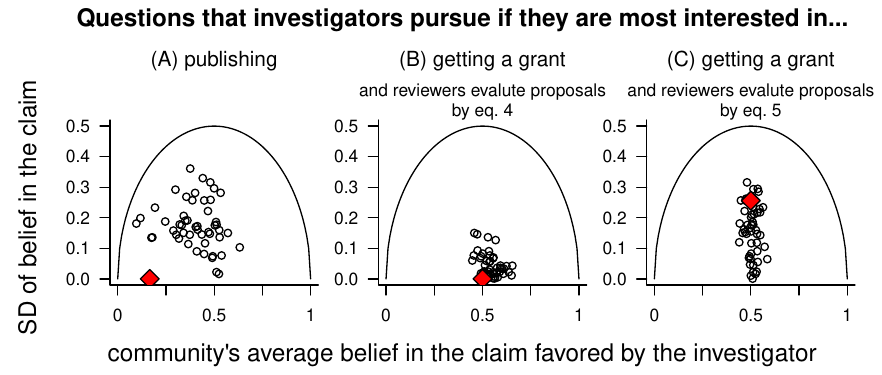}
  \end{center}
  \captionof{figure}{\textbf{Aggregate scientific activity is shaped by whether investigators are most interested in publishing or in getting a grant.}  Each panel shows the most promising questions for each of 50 investigators, where each investigator randomly samples 15 candidate questions from a large pool of available of questions and chooses to pursue the question that scores the highest with respect to one of eqs.~\ref{eq:investigator-public-word}--\ref{eq:reviewer-public-word}.  In this figure, questions are characterized by the average and standard deviation (SD) of the community's belief in the claim favored by the investigator.  Red diamonds show the optimal question for an investigator who can sample an infinite number of questions.  (A) If investigators are most interested in shifting their peers' beliefs (eq.~\ref{eq:investigator-public-word}), they pursue questions where their own beliefs diverge from those of the community, and where substantial disagreement exists within the community.  The optimal case requires a SD of beliefs infinitesimally $>0$ so that the investigator can hold the lone divergent belief. (B): If investigators are most interested in getting a grant, and if reviewers evaluate proposals based on how much they expect the experiment to change their own private beliefs (eq.~\ref{eq:reviewer-private-word}), then investigators propose questions where there is broad consensus that the competing claims are almost equally likely to be true.  (C) If investigators are most interested in getting a grant, and if reviewers evaluate proposals based on how much they expect the experiment to change the community's beliefs (eq.~\ref{eq:reviewer-public-word}), then investigators propose questions that will resolve scientific controversies, as along as each camp is not wholly set in its views.  The arc in each panel shows the upper bound on the SD of community beliefs.  See the text for how the pool of candidate questions is constructed.}
  \label{fig:best-questions}
\end{figure}

The criterion used to determine an experiment's promise has a dramatic impact on what questions members of the community pursue (Fig.~\ref{fig:best-questions}).  When investigators seek to publish (Fig.~\ref{fig:best-questions}A), they mostly pursue claims where community beliefs vary substantially, and the community on the whole is somewhat skeptical of the claim that the investigator favors.  In other words, investigators leverage the differences between their own beliefs and those of the community. The optimal question, shown by the red diamond in Fig.~\ref{fig:best-questions}A, is identical to the optimal question in the lone wolf example  (Fig.~\ref{fig:sullied-schematic}B).  However, such questions are seldom available in practice because they require that the investigator believe one claim  while everyone one else strongly believes the competing claim.   

In contrast, when investigators seek to get a grant, they instead propose questions where the community opinion on the whole is split roughly evenly between the two competing claims (Fig.~\ref{fig:best-questions}B--C).  If reviewers evaluate proposals based on their anticipated private learning (Fig.~\ref{fig:best-questions}B), then investigators are better off proposing questions in which nearly everyone in the community agrees that both claims have equal credence.  Investigators avoid proposing experiments that will resolve scientific disagreements because in those cases reviewers will find the experiments privately uninformative and will discount how the experiment may shape their peers' thinking. The optimal question in this case is one in which every community member agrees that both competing claims are equally likely.

On the other hand, if reviewers evaluate proposals based on how much they expect the experiment to shift the beliefs of the community (Fig.~\ref{fig:best-questions}C), then investigators will propose experiments that resolve scientific controversies.  In this numerical example, these controversies take the form of questions where the community is split into two opposing camps whose views roughly counterbalance.  Investigators in this example are drawn to questions that resolve mild controversies in which neither camp is so strongly set in its views as to make the disagreement intractable.  However, the strength of belief that makes that belief intractable to new evidence depends on the power of the experiment, which we have set arbitrarily here.  More powerful experiments (smaller $\sigma_Y$) would allow investigators to resolve controversies with even more radical disagreement between camps.  

In the Appendix, we provide a formal proof that the pattern observed in Fig.~\ref{fig:best-questions}B--C --- namely, that reviewers disfavor proposals that resolve controversies when they rank proposals based on the anticipated shift their own beliefs (eq.~\ref{eq:reviewer-private-word}), and favor proposals that resolve controversies when they rank proposals based on the anticipated shift their peers' beliefs (eq.~\ref{eq:reviewer-public-word}) --- holds in general for definitive experiments {\color{rcolor}that reveal the state of nature conclusively.} 

\section*{Discussion}

Dear Reader, we have written this article in hopes of shifting your beliefs --- at least somewhat --- about how the grant and publication processes shape scientific activity.   Our main point is that there is an inescapable tension between the projects that scientists favor when their primary objective is to get a grant, versus those they favor when their primary aim is to publish the experiment's results.  More broadly, this tension exists between any form of proposal-based ({\em ex ante}) vs. outcome-based ({\em ex post}) peer review.  This tension arises because investigators can leverage the differences between their private beliefs and those of the community when peer reviewers evaluate a completed experiment, but they have no opportunity to leverage these differences when peers evaluate a proposed experiment.  Of course, most scientists must both win grants and publish in order to continue in their careers, and thus these twin necessities will perpetually pull scientists in different directions.  Learning how to to navigate this tension is a key skill to succeeding in contemporary science.  Bringing it to light suggests a number of implications for the institutions that shape science.

First, because an investigator cannot leverage her own private beliefs when proposing future research, proposals for future research will inevitably be more cautious than the completed experiments reported in prestigious journals.  In our own personal experience, funders (and the scientists they enlist for review panels) are regularly disappointed by the risk-averse nature of the proposals they receive.  Our results suggest that this cautiousness is an inevitable consequence of {\em ex ante} peer review.  Investigators don't submit risky proposals because most reviewers will view risky proposals as unlikely to generate surprising results.   
This is highlighted in a recent survey of investigators who received rapid-response funding for COVID-19 research from the non-governmental ``Fast Grants'' program. There, 78\% of responding investigators said that they would change their research direction ``a lot'' with steadier, unconstrained funding at their current level, while 44\% said they would pursue hypotheses others deem unlikely \citep{collison2021what}.  

Second, there is growing momentum 
to shift publication in part towards {\em ex ante} review \citep[e.g., Registered Reports;][]{nosek2014method, nosek2018preregistration, chambers2020frontloading}.  Under this model, scientists submit designs for proposed experiments to journals, which then peer-review the design and may offer in-principle, results-agnostic acceptance for experiments deemed sufficiently meritorious.  The momentum towards {\em ex ante} journal review is intended to remedy the selective reporting, data-dredging, $p$-hacking, and other questionable research practices (QRPs) that conventional, outcome-based ({\em ex post}) review encourages \citep{sterling1959publication, munafo2017manifesto}.  Our results here suggest that {\em ex ante} journal review will screen for risk-averse experiments for the same reasons that grant proposal review does.  This is not to say that a shift toward {\em ex ante} journal review would necessarily be unwelcome, either taken alone or when weighed against the strengths and weaknesses of traditional, outcome-based review.  {\color{rcolor} Indeed, a recent analysis suggests that Registered Reports compare favorably to articles published under outcome-based review across a variety of metrics \citep{soderberg2021initial}.  However, the diminished opportunity to leverage private beliefs under {\em ex ante} review suggests that {\em ex ante} journal review would likely shift scientific activity towards a different set of questions than those pursued under {\em ex post} review.} 

Third, {\em ex ante}, reviewers will rank proposals differently if they evaluate proposals based on anticipated shifts in their own private beliefs or on anticipated shifts in the beliefs of their peers.  
When reviewers evaluate proposals based on how privately informative they expect the study to be, investigators will shy away from proposing experiments that will resolve controversies. {\color{rcolor}This leaves} the community deadlocked. On the other hand, when reviewers evaluate proposals based on their broader informativeness to the community, investigators are well served by proposing experiments that resolve controversies, as long as researchers are not so set in their views that their beliefs resist change.  In our own experience at least, funders are often silent with respect to how they expect reviewers to evaluate proposals in this respect.  We speculate that reviewers are more likely to evaluate proposals based on how their private beliefs may shift when the proposal falls further from their expertise, because a reviewer who is less familiar with the proposal's topic is also less likely to be aware of the distribution of beliefs within the relevant scholarly community.   Similarly, a reviewer who is more familiar with the science being proposed will be better positioned to assess its potential impact on the field. 
{\color{rcolor} This also suggests a motivation for convening grant reviewers as a panel. Panel deliberations reveal scientific disagreements and help reviewers assess the distribution of beliefs in the community. Thus informed, reviewers will be in a better position to identify proposals with the potential to resolve ongoing controversies.}


On the face of it, it may be surprising that a similar distinction between basing peer reviews on private versus public belief shifts does not arise {\em ex post}.  Our results here warrant further scrutiny, given the realities of outcome-based review.  If reviewers evaluate a result based only on their own private beliefs, then the average private belief shift of a small sample of reviewers may be a highly variable estimator of the {\color{rcolor} public value of the outcome.}    Of course, most journals only solicit 1 -- 3 reviews, which certainly constitutes a small sample in this regard. {\color{rcolor} In this case, luck of the reviewer draw adds a layer of stochasticity to the review process. The effect of this stochasticity on investigator behavior will depend on investigators' tolerance for uncertainty in the time and effort needed to publish in the first place.  For their part, journals can reduce this stochasticity by instructing the reviewers to evaluate manuscripts based on the aggregate belief shift the result would engender, and by selecting reviewers who can assess that shift.}

We emphasize that the analysis in this article holds most strongly when scientists contemplate pursuing questions that are roughly equivalent in other regards.  Clearly, the questions that scientists consider differ in other ways, most notably in the scientific or societal importance of the topic they examine.  When questions differ in their topical importance, a study on an important topic that will nudge beliefs by a small but non-negligible amount will likely hold more value than a study that will upend beliefs on a topic that few care about.  For example, given the high stakes, a potential and credible antiviral therapy for COVID-19 may be eminently worth testing even if the community agrees that the chance of success is small.  Yet it is worth noting that the scientific community has passed on testing some of the more outlandish COVID-19 therapies promoted by politicians, confirming that topical importance alone is not sufficient to make a study {\em ex ante} valuable if the community is sufficiently dubious about the study's potential to shift beliefs.

Finally, for the sake of simplifying matters, our set-up has intentionally ignored some other relevant complications, any of which would be interesting to explore.  First, we have portrayed an investigator's evaluation of possible experiments as a one-off process.  The great benefit to doing so is that, when investigators' beliefs diverge, we avoid having to determine which investigators' beliefs (if any) are ``correct''.  Instead, in our set-up investigators evaluate the promise of experiments based on their own beliefs, which in a one-off setting is all they can possibly do.  In real life, of course, scientists select projects sequentially, and they succeed or fail in their careers based on whether their private beliefs lead them in fruitful directions.  
Second, we have also avoided the game dynamics that may ensue when scientists compete for a limited number of grants or publication slots.  This competition may introduce strategic considerations into an investigator's project choice in ways that our current set-up does not accommodate.  {\color{rcolor} Third, while it is useful to juxtapose the private and public value of belief shifts, human psychology is likely to lead researchers and reviewers to value changes in their own vs.\ others' beliefs differently based on context.  For example, when controversies run hot, reviewers may favor projects or outcomes that validate their own beliefs and disfavor projects our outcomes that do the opposite \citep{latour1987science}.}
Finally, and of course, scientists' actual beliefs are not atomized into a collection of isolated beliefs for individual claims, but instead are part of each individual scientists' larger conception of the natural world.  A richer model for how scientists' interconnected beliefs evolve would provide a valuable contribution to the study of science indeed.

\section*{Acknowledgments}
We thank Jacob Foster, Ryan Martin, Mark Tanaka, and Kevin Zollman for helpful discussions.  We thank NSF for supporting this work through awards SMA 19-52343 to KG and SMA 19-52069 to CTB.  KG also thanks the University of Washington Department of Biology for visitor support.

\bibliography{filters}

\renewcommand\thefigure{A\arabic{figure}}  
\setcounter{figure}{0}  

\renewcommand\theequation{A.\arabic{equation}}  
\setcounter{equation}{0}   

\section*{Appendix}

\subsection*{Generalization to any valid measure of information}

Frankel \& Kamenica \citep{frankel2019quantifying} prove that any ``valid'' measure of information that maps a Bayesian agent's prior and posterior beliefs to a real number can be microfounded in some (not necessarily unique) decision problem.   In the main text, we focused on the case where the investigator's utility was determined by how accurately her beliefs anticipated the state of nature, or observable phenomena that derive from it.  Here, we use Frankel \& Kamenica's arguments to extend our model to any utility function that an investigator may face.

Much of what follows is adapted directly from reference \cite{frankel2019quantifying}.  Let $A$ be an action space with typical element $a$, and let $u(a, x)$ give the observer's utility obtained by taking action $a$ when the true state of nature is $x$.  A pair $(A, u)$ defines a decision problem in this context.  

Let {\color{rcolor}$a^*(P) \in \argmax_{a \in A} \int u(a,x) \, dP(x)$} be a utility-optimizing action to a risk-neutral decision-maker who believes ${\color{rcolor} P}$.  Now suppose this decision maker holds prior belief ${\color{rcolor} P}$, observes $Y=y$, and uses Bayesian reasoning to update her beliefs to ${\color{rcolor} Q}$.  For a given decision problem, the instrumental value of observing $y$ to this observer is 
\begin{equation}
    {\color{rcolor} \mathcal{V}(Q, P) = \int \, \left( u(a^*(Q), x) - u(a^*(P), x) \right) \, dQ(x)}.
    \label{eq:fk-value}
\end{equation}
In other words, the value is the amount by which the decision maker perceives her expected utility to have increased by updating her beliefs from ${\color{rcolor} P}$ to ${\color{rcolor} Q}$.  The quantity ${\color{rcolor} \mathcal{V}(Q, P)}$ is the counterpart to ${\color{rcolor} d(Q || P)}$  in the main text, and generalizes our set-up accordingly. 

A decision problem that gives rise to the setting described in the main text is one in which the action $A \in {\color{rcolor} \mathcal{P}}$ is an announcement of a set of beliefs, and the utility is $u(a,x) = S(\delta_x, x) - S(a,x)$, where $\delta_x$ is the degenerate distribution that assigns probability 1 to $X=x$.  Thus, the investigator maximizes her expected utility by minimizing the expected score of her beliefs.


\subsection*{Modes of {\em ex ante} review and scientific controversies}

Here, we argue that the distinction between eq.~\ref{eq:reviewer-private-word} and eq.~\ref{eq:reviewer-public-word} that appears in Fig.~\ref{fig:best-questions}B--C follows from the properties of proper scoring rules, or of any valid measure of information (sensu ref.\ \citep{frankel2019quantifying}) more generally.  The results that we establish below are limited to definitive experiments, that is, experiments that reveal the state of nature $X$ conclusively. 

{\color{rcolor} Let $\bar{P} = \int P \, d\Pi(P)$}
be the {\color{rcolor} average belief held within a} community.  We show the following.  First, for a given ${\color{rcolor} \bar{P}}$, if reviewers rate proposals for definitive experiments based on the expected shift in their own belief (eq.~\ref{eq:reviewer-private-word}), then any heterogeneity in beliefs among reviewers will lower the proposal's value relative to the case in which all reviewers hold the same beliefs.  Second, for a given ${\color{rcolor} \bar{P}}$, if reviewers rate proposals for definitive experiments based on the expected shift in their peers' beliefs (eq.~\ref{eq:reviewer-public-word}), then any heterogeneity in beliefs among reviewers will raise the proposal's value relative to the case in which all reviewers hold the same beliefs.  These results hold strictly for strictly proper scoring rules, and they hold weakly for any valid measure of information as defined in reference \citep{frankel2019quantifying}.

Our proof requires the concept of a generalized entropy, defined as the expected value of a scoring rule $S({\color{rcolor} P},x)$ when $x$ is distributed according to ${\color{rcolor} P}$:
\begin{equation}
    {\color{rcolor} e(P) = \int S(P, x) \ dP(x)}.
    \label{eq:entropy}
\end{equation}
(Plugging the ignorance score into eq.~\ref{eq:entropy} gives the familiar Shannon entropy from information theory.)  Strict propriety of the scoring rule implies strict concavity of the entropy \citep{brocker2009reliability, gneiting2007strictly}.

The first part of the result --- that heterogeneity in reviewer beliefs decreases the average review score under eq.~\ref{eq:reviewer-private-word} --- follows immediately from the concavity of $e({\color{rcolor} P})$.  Namely, in a definitive experiment, each reviewer's expected change in beliefs equals the entropy of their beliefs.  Then Jensen's inequality establishes {\color{rcolor}$\int e(P) \, d\Pi(P) < e(\bar{P})$ for any non-degenerate $\Pi$}. 

The second part of the result requires a bit more machinery.  For a particular scoring rule $S$, define the associated scoring function $\mathbb{S}({\color{rcolor} R}, {\color{rcolor} P})$ as 
\begin{equation}
    {\color{rcolor}\mathbb{S}(R, P) = \int S(R, x) \ dP(x)}.
    \label{eq:scoring-function}
\end{equation}
In other words, the scoring function $\mathbb{S}({\color{rcolor} R}, {\color{rcolor} P})$ is the expectation of the scoring rule when the forecast ${\color{rcolor} R}$ is issued and $x$ is distributed according to ${\color{rcolor} P}$.  For scoring rules, strict propriety of the scoring rule $S$ implies \citep{brocker2009reliability}
\begin{equation}
    {\color{rcolor} e(P) = \inf_{R \in \mathcal{P}} \mathbb{S}(R, P)}.
\end{equation}
For definitive experiments, the average amount that a reviewer expects the community to learn (eq.~\ref{eq:reviewer-public-word}) is 
{\color{rcolor}
\begin{equation}
    \int \! \int \mathbb{S}(P,R) \, d\Pi(R) \, d\Pi(P).
    \label{eq:reviewer-public-appendix}
\end{equation}}
Because $\mathbb{S}$ is linear in its second argument, eq.~\ref{eq:reviewer-public-appendix} equals {\color{rcolor} $\int \mathbb{S}(P,\bar{P}) \, d\Pi(P)$}.  However, for a given ${\color{rcolor}\bar{P}}$, the scoring function {\color{rcolor} $\mathbb{S}(P,\bar{P})$} is minimized when {\color{rcolor} $P = \bar{P}$}.  Hence {\color{rcolor} $\int \mathbb{S}(P,\bar{P}) \, d\Pi(P)$} achieves its minimum when ${\color{rcolor} \Pi}$ takes a degenerate distribution at {\color{rcolor}$\bar{P}$}.

To extend these results to any utility function, replace eq.~\ref{eq:entropy} with 
\begin{equation}
    {\color{rcolor} c(P) = \int \, \left( u(a^*(\delta_x), x) - u(a^*(P), x) \right) \, dP(x)}
    \label{eq:uncertainty}
\end{equation}
and replace eq.~\ref{eq:scoring-function} with
\begin{equation}
    {\color{rcolor} \mathbb{C}(R,P) = \int \, \left( u(a^*(\delta_x), x) - u(a^*(R), x) \right) \, dP(x)}.
    \label{eq:scoring-function-generalized}
\end{equation}
Eq.~\ref{eq:uncertainty} is the measure of uncertainty coupled to \ref{eq:fk-value}, and it equals the utility loss that a decision-maker with belief {\color{rcolor}$P$} expects from not knowing $x$ exactly \cite{frankel2019quantifying}. Eq.~\ref{eq:scoring-function-generalized} is the utility loss of uncertainty to someone who believes {\color{rcolor}$R$} as assessed by an observer who believes {\color{rcolor}$P$}.  For some decision problems \ref{eq:uncertainty} may only be weakly concave in {\color{rcolor}$P$}.  Thus, the proof above can be extended by replacing $e({\color{rcolor}P})$ with $c({\color{rcolor}P})$, replacing $\mathbb{S}({\color{rcolor}R},{\color{rcolor}P})$ with $\mathbb{C}({\color{rcolor}R},{\color{rcolor}P})$, and replacing strict inequalities with weak ones.

\end{document}